\documentclass[aps,twocolumn,a4paper,showpacs,nofootinbib,superscriptaddress]{revtex4}
\usepackage[dvips]{graphicx}
\usepackage{amssymb}

\begin{document}

\title{Generalized virial theorem for massless electrons in graphene\\ and other Dirac materials}
\author{A.A. Sokolik}%
\affiliation{Institute for Spectroscopy, Russian Academy of Sciences, 142190 Troitsk, Moscow, Russia}%
\affiliation{MIEM at National Research University HSE, 109028 Moscow, Russia}%
\author{A.D. Zabolotskiy}%
\affiliation{Institute for Spectroscopy, Russian Academy of Sciences, 142190 Troitsk, Moscow, Russia}%
\affiliation{Dukhov Research Institute of Automatics (VNIIA), 127055 Moscow, Russia}%
\author{Yu.E. Lozovik}%
\email{lozovik@isan.troitsk.ru}
\affiliation{Institute for Spectroscopy, Russian Academy of Sciences, 142190 Troitsk, Moscow, Russia}%
\affiliation{MIEM at National Research University HSE, 109028 Moscow, Russia}
\affiliation{Dukhov Research Institute of Automatics (VNIIA), 127055 Moscow, Russia}%
\affiliation{Moscow Institute of Physics and Technology, 141700 Dolgoprudny, Moscow region, Russia}%
\begin{abstract}
The virial theorem for a system of interacting electrons in a crystal, which is described within the framework of the
tight-binding model, is derived. We show that, in particular case of interacting massless electrons in graphene and
other Dirac materials, the conventional virial theorem is violated. Starting from the tight-binding model, we derive
the generalized virial theorem for Dirac electron systems, which contains an additional term associated with a momentum
cutoff at the bottom of the energy band. Additionally, we derive the generalized virial theorem within the Dirac model
using the minimization of the variational energy. The obtained theorem is illustrated by many-body calculations of the
ground state energy of an electron gas in graphene carried out in Hartree-Fock and self-consistent random-phase
approximations. Experimental verification of the theorem in the case of graphene is discussed.
\end{abstract}


\maketitle

\section{Introduction}

The virial theorem (VT) provides an exact relationship between mean kinetic and potential energies of classical and
quantum systems when these energies are power-law functions of coordinates and momenta of the constituent particles
\cite{Marc,Weislinger}. Applications of VT include estimations of various quantities in physics of atoms, molecules,
solids, plasmas, and in astrophysics. Additionally, VT and its hypervirial generalizations are used to calculate matrix
elements, to improve variational calculations in quantum chemistry, as well as to check and improve equations of states
and electron density functionals (see \cite{Marc,Weislinger,Vasilev,Levy,Parr} and references therein).

From the fundamental point of view, VT is associated with the scale transformations of the system \cite{Marc,Gaite} or
with dilatations of a ground state wave function in the quantum-mechanical case \cite{Fock,Lowdin}. When a spatially
confined many-particle system (for example, electron gas in a finite-sized solid) is considered, VT acquires an
additional ``boundary'' term proportional to the thermodynamic pressure in the system \cite{Marc}, that can be
interpreted as a manifestation of non-Hermiticity of the Hamiltonian \cite{Abad,Esteve}. The same term appears in VT
for a uniform electron gas, that can be attributed to existence of an intrinsic length scale in the system, namely, the
Bohr radius \cite{March1,Argyres}.

There is a number of materials intensively studied in recent years where electrons at low excitation energies behave as
massless particles: two-dimensional graphene \cite{CastroNeto,Kotov}, surfaces of topological insulators
\cite{Hasan,Qi}, three-dimensional Dirac semimetals \cite{Wehling}, and Weyl semimetals, which were discovered in very
recent experiments \cite{Lv,Xu,Yang}. When Dirac electrons are massive, their kinetic energy is not a power-law
function of momentum and hence VT does not provide exact relationship between kinetic and potential energies
\cite{Fock,Rose,March2,Bachall}. However for massless Dirac electrons with linear dispersion such relationship can be
obtained.

In this article, we derive the virial theorem for Coulomb-interacting massless electrons in Dirac materials. First, on
the example of graphene, we obtain general VT for electrons in a crystal, described by a tight-binding model. Then we
proceed to the approximate model of massless Dirac electrons with the momentum cutoff imposed at the bottom of the
valence band. The obtained theorem should be as accurate as the Dirac approximation for electron dynamics.

It is shown that massless electrons in a Dirac material disobey conventional VT but instead they can be described by
the generalized virial theorem (GVT), which contains an additional term proportional to the derivative of the ground
state energy with respect to the cutoff momentum. This term can be interpreted as a manifestation of the underlying
crystal lattice, characterized by a definite lattice constant, during the scale transformations. The role of a periodic
potential of a crystal lattice in VT for electrons was also studied in the Kronig-Penney model \cite{Weislinger2} and
in the density functional context \cite{Mirhosseini}. Another approach to obtain GVT presented in the article is the
restricted minimization of the variational energy with respect to dilatations of the ground state wave function.

The obtained GVT is illustrated in the case of graphene by the many-body diagrammatic calculations in Hartree-Fock and
self-consistent random-phase approximations. It is shown that the relative contribution of the cutoff-induced term in
GVT for graphene ranges from 5 to 20\% at typical conditions. However, the obtained form of GVT is applicable to any
other Dirac material with massless electrons near the Fermi energy.

The article is organized as follows. We derive GVT for graphene in the tight-binding model in Sec.~II and obtain its
approximation in the Dirac model in Sec.~III. GVT for Dirac materials will also be obtained in Sec.~IV by means of
restricted minimization. We illustrate this theorem by many-body calculations of the ground state properties of the
electron gas in graphene in Sec.~V, considering separately the cutoff-induced term in Sec.~VI, and make conclusions in
Sec.~VII.

\section{Virial theorem in the tight-binding model}

The many-body Hamiltonian of $2\mathrm{p}_z$-electrons in graphene $H=H_0+H_\mathrm{int}+H_\mathrm{ext}$ includes the
tight-binding nearest-neighbor kinetic energies
\begin{eqnarray}
H_0=\sum_{i}\left(\begin{array}{cc}\epsilon_0&-tf_{\mathbf{p}_i}\\-tf^*_{\mathbf{p}_i}&\epsilon_0\end{array}\right)_i,
\label{H_0}
\end{eqnarray}
as well as Coulomb electron-electron interaction and interactions of electrons with the external potential
$U_\mathrm{ext}(\mathbf{r})$:
\begin{eqnarray}
H_\mathrm{int}=\frac12\sum_{i\neq j}\frac{e^2}{\varepsilon|\mathbf{r}_i-\mathbf{r}_j|},\qquad
H_\mathrm{ext}=\sum_{i}U_\mathrm{ext}(\mathbf{r}_i).
\end{eqnarray}
Here $\epsilon_0$ and $t$ are the on-site energy and the hopping integral,
$f_\mathbf{p}=\sum_{s=1}^3e^{(i/\hbar)\mathbf{p}\mathbf{b}_s}$, where $\mathbf{b}_s$ are the vectors, which connect any
atom from graphene sublattice $A$ with its nearest neighbors from sublattice $B$ \cite{CastroNeto,Kotov}, $\varepsilon$
is the dielectric constant of background medium; the matrix in (\ref{H_0}) acts on sublattice degree of freedom of the
$i$-th electron.

As the starting point for derivation of the virial theorem, we use, similarly to \cite{Abad,Esteve}, the following
equation:
\begin{eqnarray}
\langle\Psi|[H,G]|\Psi\rangle=0,\label{virial_general}
\end{eqnarray}
which is satisfied identically for any stationary state $|\Psi\rangle$ and for any operator $G$, provided the
Hamiltonian $H$ is Hermitian. To obtain VT, $G$ is chosen to be the virial operator:
\begin{eqnarray}
G=\sum_i\frac{\mathbf{r}_i\mathbf{p}_i+\mathbf{p}_i\mathbf{r}_i}2.\label{G}
\end{eqnarray}
The commutators of each term of $H$ with $G$ can be calculated explicitly:
\begin{eqnarray}
[H_0,G]=-i\hbar\sum_i\mathbf{p}_i\frac{\partial H_0}{\partial\mathbf{p}_i},\quad[H_\mathrm{int},G]=-i\hbar
H_\mathrm{int},\label{G_commutators}
\end{eqnarray}
\begin{eqnarray}
[H_\mathrm{ext},G]=i\hbar\sum_i\mathbf{r}_i\frac{\partial
U_\mathrm{ext}(\mathbf{r}_i)}{\partial\mathbf{r}_i}.\label{H_ext_commutator1}
\end{eqnarray}

Now we apply the general formulas (\ref{virial_general})--(\ref{H_ext_commutator1}) to the tight-binding model, where
the electron momenta $\mathbf{p}_i$ enter (\ref{H_0}) only in a product with the lattice constant $a$ (since
$|\mathbf{b}_s|=a/\sqrt3$), thus we can write
\begin{eqnarray}
[H_0,G]=-i\hbar a\left(\frac{\partial H_0}{\partial a}\right)_{\epsilon_0,t}.\label{H_0_commutator1}
\end{eqnarray}
The derivative with respect to $a$ is taken here at $\epsilon_0,t=\mathrm{const}$ because otherwise we would obtain
unnecessary terms, which arise due to the physical dependence of $\epsilon_0$ and $t$ on $a$.

For the external potential $U_\mathrm{ext}(\mathbf{r})$, which confines electrons within a definite spatial volume, we
can write: $U_\mathrm{ext}(\mathbf{r})=\tilde{U}(\mathbf{r}/R)$, where $R$ is the linear size of this volume. Using
this relation in (\ref{H_ext_commutator1}), we get:
\begin{eqnarray}
[H_\mathrm{ext},G]=-i\hbar R\frac{\partial H_\mathrm{ext}}{\partial R}.\label{H_ext_commutator2}
\end{eqnarray}
Substituting (\ref{G_commutators}), (\ref{H_0_commutator1}), (\ref{H_ext_commutator2}) to (\ref{virial_general}) and
using the Hellmann-Feynman theorem, we obtain:
\begin{eqnarray}
-a\left(\frac{\partial\tilde{E}}{\partial a}\right)_{\epsilon_0,t}-\tilde{E}_\mathrm{int} -R\frac{\partial
\tilde{E}}{\partial R}=0,\label{virial_tb1}
\end{eqnarray}
where $\tilde{E}=\langle\Psi|H|\Psi\rangle$ and $\tilde{E}_\mathrm{int}=\langle\Psi|H_\mathrm{int}|\Psi\rangle$ are,
respectively, tight-binding ground state and interaction energies.

Eq.~(\ref{virial_tb1}) is the tight-binding version of VT. It is derived explicitly for graphene, but can be applied
for any tight-binding model with the condition that all hopping integrals and on-site energies are constant upon taking
the derivative $\partial\tilde{E}/\partial a$.

\begin{figure}[t]
\begin{center}
\resizebox{0.65\columnwidth}{!}{\includegraphics{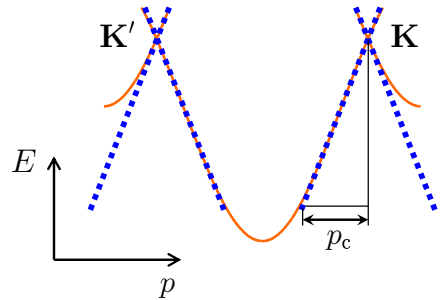}}
\end{center}
\caption{\label{Fig1}(Color online) Schematic tight-binding dispersion of filled single-electron states in
electron-doped graphene (solid line) and its approximation in the Dirac model (dotted line). The electron states in the
Dirac model are filled in both valleys $\mathbf{K}$ and $\mathbf{K}'$ down to the cutoff momentum $p_\mathrm{c}$.}
\end{figure}

\section{Virial theorem for massless Dirac electrons}

When the Fermi energy in graphene (or any other Dirac material) is close to the Dirac point, it is a reasonable
approximation to switch from the tight-binding model of single-electron dynamics to the effective Dirac model
\cite{CastroNeto,Kotov,Wehling} (Fig.~\ref{Fig1}). In this model, the momentum cutoff $p_\mathrm{c}$ in the valence
band should be introduced in order to ensure that the total density of electrons is equal to their actual density in
the crystal.

The Hamiltonian of interacting massless Dirac electrons is $H=H_\mathrm{D}+H_\mathrm{int}+H_\mathrm{ext}$, where
\begin{eqnarray}
H_\mathrm{D}=v_\mathrm{F}\sum_{ic}\left(\begin{array}{cc}0&cp_x-ip_y\\cp_x+ip_y&0\end{array}\right)_{ic} \label{H_D}
\end{eqnarray}
is the Dirac Hamiltonian of noninteracting electrons of chiralities $c=\pm1$ and $v_\mathrm{F}=at\sqrt3/2\hbar$ is the
Fermi velocity. In this approach, the total number $\tilde{N}=N_0+N$ of electrons that fill the energy band is a sum of
the number $N_0\propto(R/a)^D$ of electrons, that fill the valence band from the cutoff momentum $p_\mathrm{c}$ at its
bottom up to the Dirac point ($D$ is the space dimension), and the much smaller number $N$ of Dirac carriers (electrons
at $N>0$ or holes at $N<0$). Similarly we can define the full energy of Dirac electrons
\begin{eqnarray}
E(N)=\tilde{E}(\tilde{N})-\tilde{E}(N_0)-\epsilon_0N\label{Dirac_energy}
\end{eqnarray}
[the last term redefines the on-site energy in the Dirac single-electron Hamiltonian (\ref{H_D}) to zero] and their
interaction energy $E_\mathrm{int}(N)=\tilde{E}_\mathrm{int}(\tilde{N})-\tilde{E}_\mathrm{int}(N_0)$. Note that these
energies also include the energy of interaction between the $N$ Dirac electrons and the $N_0$ electrons of the filled
valence band. Subtracting from (\ref{virial_tb1}) the same equation at $\tilde{N}=N_0$, we get:
\begin{eqnarray}
-a\left(\frac{\partial {E}}{\partial a}\right)_{\tilde{N}}-E_\mathrm{int} -R\left(\frac{\partial E}{\partial
R}\right)_{\tilde{N}}=0.\label{virial_tb2}
\end{eqnarray}
Hereafter we assume $\epsilon_0,t=\mathrm{const}$ in all derivatives; we have also showed explicitly the condition
$\tilde{N}=\mathrm{const}$ implied by the Hellmann-Feynman theorem for the $\tilde{N}$-particle ground state wave
function.

In the Dirac model for massless electrons, the electron gas properties depend on the lattice constant $a$ indirectly
via $v_\mathrm{F}\propto a$ and the cutoff momentum $p_\mathrm{c}\propto a^{-1}$ \cite{CastroNeto,Kotov}, so that the
energy $E$ can be calculated as a function $E(N,v_\mathrm{F},p_\mathrm{c},R)$, where $N=\tilde{N}-N_0$. Consequently,
the derivatives of $E$ in (\ref{virial_tb2}) with taking into account (\ref{Dirac_energy}) are:
\begin{eqnarray}
a\!\left(\frac{\partial {E}}{\partial a}\right)_{\!\!\tilde{N}}\!\!=-\!\left(\frac{\partial E}{\partial
N}+\epsilon_0\right)\!a\frac{\partial N_0}{\partial a}+v_\mathrm{F}\frac{\partial E}{\partial
v_\mathrm{F}}-p_\mathrm{c}\frac{\partial E}{\partial p_\mathrm{c}},
\label{der1}\\
R\!\left(\frac{\partial {E}}{\partial R}\right)_{\!\!\tilde{N}}\!\!=-\!\left(\frac{\partial E}{\partial
N}+\epsilon_0\right)\!R\frac{\partial N_0}{\partial R}+R\frac{\partial E}{\partial R}.\label{der2}
\end{eqnarray}
Since $N_0\propto(R/a)^D$, the first terms in right-hand sides of (\ref{der1}) and (\ref{der2}) cancel upon
substituting to (\ref{virial_tb2}). Additionally, using the Hellmann-Feynman theorem for the kinetic energy
$E_\mathrm{kin}=\langle\Psi|H_\mathrm{D}|\Psi\rangle=v_\mathrm{F}(\partial E/\partial v_\mathrm{F})$, we get the
following form of GVT for massless Dirac particles:
\begin{eqnarray}
-E-R\frac{\partial E}{\partial R}+p_\mathrm{c}\frac{\partial E}{\partial p_\mathrm{c}}=0,\label{virial_Dirac1}
\end{eqnarray}
where $N=\mathrm{const}$ during differentiations. The specific character of electrons with linear dispersion and
Coulomb interaction is revealed in the fact that kinetic and interaction energies enter VT with the same coefficients,
yielding the full energy $E=E_\mathrm{kin}+E_\mathrm{int}$ \cite{Footnote1}; that reflects the absence of intrinsic
length parameter in the system. The only length parameter is introduced by the crystal lattice via the cutoff momentum,
$p_\mathrm{c}$, which provides the last term in GVT (\ref{virial_Dirac1}).

The second term in (\ref{virial_Dirac1}) can be related to the pressure $p=-(\partial E/\partial V)_N$ of the electron
gas,
\begin{eqnarray}
-R\frac{\partial E}{\partial R}=DpV,
\end{eqnarray}
where $V\propto R^D$ is the $D$-dimensional volume of the system.

In the case of a homogeneous electron gas, assuming extensiveness of the energy $E=R^D\epsilon(N/R^D)$ with respect to
the $D$-dimensional volume $V\propto R^D$, we have:
\begin{eqnarray}
R\frac{\partial E}{\partial R}=DE-DN\frac{\partial E}{\partial N},\label{dE_dR}
\end{eqnarray}
therefore GVT (\ref{virial_Dirac1}) takes the form:
\begin{eqnarray}
-(D+1)E+DN\frac{\partial E}{\partial N}+p_\mathrm{c}\frac{\partial E}{\partial p_\mathrm{c}}=0,\label{virial_Dirac3}
\end{eqnarray}
where $R=\mathrm{const}$.

For some calculations in the limit $T\rightarrow0$, it is more convenient to switch to the grand canonical ensemble,
where the state of the system is characterized by the (zero-temperature) grand thermodynamic potential $\Omega=E-\mu
N$. The Legendre transformation
\begin{eqnarray}
E(N,p_\mathrm{c})=\Omega(\mu,p_\mathrm{c})+\mu N
\end{eqnarray}
implies the relations $\partial E/\partial N=\mu$, $\partial\Omega/\partial\mu=-N$, $\partial E/\partial
p_\mathrm{c}=\partial\Omega/\partial p_\mathrm{c}$, that, upon substituting in (\ref{virial_Dirac3}), provide GVT in
terms of $\Omega$:
\begin{eqnarray}
-(D+1)\Omega+\mu\frac{\partial\Omega}{\partial\mu}+p_\mathrm{c}\frac{\partial\Omega}{\partial p_\mathrm{c}}=0.
\label{virial_Dirac2}
\end{eqnarray}
Extensiveness of the thermodynamic potential allows to relate it to the pressure: $\Omega=-pV$.

The chemical potential $\mu$ in Dirac electron systems is usually measured from the Dirac point, which implies $\mu=0$
at $N=0$ (with regard to graphene it is a charge neutrality point). However, generally it is not the case for a system
of interacting electrons, i.e. the ``background'' chemical potential
\begin{eqnarray}
\mu_0=\left.\frac{\partial E}{\partial N}\right|_{N=0}\label{mu_0}
\end{eqnarray}
is not zero. For example, in the case of graphene,
\begin{eqnarray}
\mu_0=-\frac{e^2p_\mathrm{c}}{2\varepsilon\hbar}\label{mu_0_graphene}
\end{eqnarray}
in the Hartree-Fock approximation \cite{Hwang}.

In Appendix \ref{Appendix_A}, the energy $E$ and the thermodynamic potential $\Omega$ are regularized in order to
remove the contributions of $\mu_0$. The virial theorems (\ref{virial_Dirac4}), (\ref{virial_Dirac5}), expressed in
terms of these regularized quantities, are derived and shown to have the same form as (\ref{virial_Dirac3}),
(\ref{virial_Dirac2}) due to the relation $\mu_0\propto p_\mathrm{c}$.

\section{Restricted minimization}

GVT for Dirac electrons (\ref{virial_Dirac1}) can be obtained directly by minimizing the energy with respect to
dilatations of the ground state wave function, similarly to its original variational derivation in \cite{Fock,Lowdin}.
Let
\begin{eqnarray}
\Psi_\lambda(\mathbf{r}_1\ldots\mathbf{r}_N)=
\frac1{\lambda^{ND/2}}\Psi\left(\frac{\mathbf{r}_1}\lambda\ldots\frac{\mathbf{r}_N}\lambda\right)\label{Dilatation}
\end{eqnarray}
be the ground state wave function $\Psi$, which is stretched by a factor of $\lambda$ uniformly in all directions.
Without resorting to the coordinate representation, we can write $|\Psi_\lambda\rangle=D_\lambda|\Psi\rangle$, where
\begin{eqnarray}
D_\lambda=\exp\left\{-\frac{i}\hbar G\ln\lambda\right\}\label{D}
\end{eqnarray}
is the dilatation operator \cite{Abad,Carinena} and $G$ is given by (\ref{G}).

Since the energies of Dirac particles are unbounded from below, we cannot minimize
$\langle\Psi_\lambda|H|\Psi_\lambda\rangle$ to obtain the ground state. One of the possible ways to overcome this
obstacle, which is most appropriate to the physical problem of a Dirac electron gas in a solid (Fig.~\ref{Fig1}), is to
impose the momentum cutoff, requiring that all electron momenta in the many-body state $|\Psi\rangle$ do not exceed
$p_\mathrm{c}$. This can be achieved by using the operator $P_{p_\mathrm{c}}$ which projects the many-body state vector
on a subspace of Slater determinants, that consists of all possible single-particle plane-wave states with momenta
$|\mathbf{p}|\leq p_\mathrm{c}$ (see Appendix \ref{Appendix_B}).

Although the ground state $|\Psi\rangle$ belongs to the subspace of $P_{p_\mathrm{c}}$, i.e.
$P_{p_\mathrm{c}}|\Psi\rangle=|\Psi\rangle$, the dilated state $|\Psi_\lambda\rangle$ generally does not belong to this
subspace. Therefore, in our restricted minimization procedure, we need to return $|\Psi_\lambda\rangle$ back into the
subspace at every $\lambda$, acting on it by $P_{p_\mathrm{c}}$ and normalizing the resulting variational state vector.
The variational energy
\begin{eqnarray}
E_\lambda=\frac{\langle\Psi|D_\lambda^{-1}P_{p_\mathrm{c}}HP_{p_\mathrm{c}}D_\lambda|\Psi\rangle}
{\langle\Psi|D_\lambda^{-1}P_{p_\mathrm{c}}D_\lambda|\Psi\rangle},\label{min1}
\end{eqnarray}
which is obtained via this restriction procedure, must have a minimum in the ground state, when $\lambda=1$, thus its
derivative in $\lambda$ must vanish:
\begin{eqnarray}
\left.\frac{\partial E_\lambda}{\partial\lambda}\right|_{\lambda=1}\!\!\!\!=
-\frac{i}\hbar\langle\Psi|[P_{p_\mathrm{c}}HP_{p_\mathrm{c}},G]-E[P_{p_\mathrm{c}},G]|\Psi\rangle=0.\label{min2}
\end{eqnarray}
Here we used (\ref{D}) and denoted $E=E_{\lambda=1}$.

The second term in (\ref{min2}) vanishes because $P_{p_\mathrm{c}}|\Psi\rangle=|\Psi\rangle$ for the ground state. The
first term, with a help of (\ref{P_pc_comm}), can be presented as
\begin{eqnarray}
[P_{p_\mathrm{c}}HP_{p_\mathrm{c}},G]=P_{p_\mathrm{c}}[H,G]P_{p_\mathrm{c}}+i\hbar p_\mathrm{c}\frac\partial{\partial
p_\mathrm{c}}(P_{p_\mathrm{c}}HP_{p_\mathrm{c}}).
\end{eqnarray}
Using the Hellmann-Feynman theorem, Eq.~(\ref{min2}) will result in
\begin{eqnarray}
\left.\frac{\partial E_\lambda}{\partial\lambda}\right|_{\lambda=1}\!\!\!\!=
-\frac{i}\hbar\langle\Psi|[H,G]|\Psi\rangle+p_\mathrm{c}\frac{\partial E}{\partial p_\mathrm{c}}=0.\label{min3}
\end{eqnarray}
Comparing (\ref{min3}) with (\ref{virial_general}), we see that the restriction, which is imposed on the state vector
during minimization, results in appearance of an additional term in GVT.

For Dirac electrons with the Hamiltonian $H=H_\mathrm{D}+H_\mathrm{int}+H_\mathrm{ext}$, similarly to
(\ref{G_commutators}), (\ref{H_ext_commutator2}), we have
\begin{eqnarray}
[H,G]=i\hbar\left(-H_\mathrm{D}-H_\mathrm{int}-R\frac{\partial H_\mathrm{ext}}{\partial R}\right).\label{comm2}
\end{eqnarray}
Substituting (\ref{comm2}) and neglecting $\langle\Psi|H_\mathrm{ext}|\Psi\rangle$ (see \cite{Footnote1}), we get the
same GVT for Dirac particles (\ref{virial_Dirac1}) as obtained in Sec.~III from the tight-binding model.

Alternatively, instead of imposing the hard cutoff $|\mathbf{p}|<p_\mathrm{c}$, we can supplement the Hamiltonian with
the term
\begin{eqnarray}
\Delta H=(-H_\mathrm{D}+A)(1-P_{p_\mathrm{c}}),\label{Delta_H}
\end{eqnarray}
which ensures that the electron states with $|\mathbf{p}|>p_\mathrm{c}$ have large positive energy $A$ and so pushes
the ground state $|\Psi\rangle$ into the subspace of $P_{p_\mathrm{c}}$. In this case the system energy becomes bounded
from below and the ground state $|\Psi\rangle$ becomes well-defined. Therefore we can apply the conventional VT
(\ref{virial_general}) with $H+\Delta H$ instead of $H$. Using (\ref{comm2}) and the property (\ref{P_pc_comm}), we
get:
\begin{eqnarray}
\left\langle\Psi\left|-H_\mathrm{D}-H_\mathrm{int}-R\frac{\partial H_\mathrm{ext}}{\partial
R}-Ap_\mathrm{c}\frac{\partial P_{p_\mathrm{c}}}{\partial p_\mathrm{c}}\right|\Psi\right\rangle\nonumber\\
+\langle\Psi|H_0(1-P_{p_\mathrm{c}})|\Psi\rangle=0.
\end{eqnarray}
The last term is proportional to the part $(1-P_{p_\mathrm{c}})|\Psi\rangle$ of the ground state vector, which goes
beyond the low-energy subspace of $|\Psi\rangle$ and thus behaves as $\sim1/A$ at $A\rightarrow\infty$. Then if we
apply the Hellmann-Feynman theorem to $\partial E/\partial R$ and $\partial E/\partial p_\mathrm{c}$, and tend to the
limit $A\rightarrow\infty$, we again obtain GVT (\ref{virial_Dirac1}).

This last method does not require $P_{p_\mathrm{c}}|\Psi\rangle=|\Psi\rangle$ in the ground state, which is, generally,
not the case in nonuniform systems where $H$ and $P_{p_\mathrm{c}}$ do not commute.

\section{Many-body calculations for graphene}

In the case of graphene, we can illustrate GVT (\ref{virial_Dirac1}) with many-body calculations of the ground state
energy $E$. Throughout Secs. V and VI we assume $D=2$ and use the symbol $S$ instead of $V$ for a two-dimensional
volume (area).

As known, the interaction-induced correction $\delta\Omega=\Omega-\Omega_0$ to the grand thermodynamic potential
$\Omega(\mu)$ at given chemical potential $\mu$ can be calculated as a sum of closed connected diagrams
\cite{Abrikosov,Mahan}; $\Omega_0(\mu)$ is the thermodynamic potential of the noninteracting electron gas. Summation of
infinite diagrammatic series is usually carried out by means of Dyson equations, where bare Green functions are
replaced with the dressed ones; thus the diagrammatic calculations become self-consistent. Analogous replacements in
closed diagrams for $\delta\Omega$ produce excess terms, which, however, can be compensated by using the Luttinger-Ward
functional \cite{Luttinger} (see its application for graphene in \cite{Lozovik}).

Here we use another approach, also described in \cite{Luttinger}: at $T=0$ in the spherically-symmetric system,
self-consistent calculation of the ground state energy $E(N)$ at given electron number $N$ can be carried out according
to the formula
\begin{eqnarray}
E(N)=\Omega_0(\mu_0(N))+\mu_0(N)N+\delta\Omega_\mathrm{BG}(\mu_0(N)),\label{BG}
\end{eqnarray}
where $\mu_0(N)$ is the chemical potential of a noninteracting electron gas at given $N$ and
$\delta\Omega_\mathrm{BG}(\mu)$ is the Brueckner-Goldstone sum of all closed connected diagrams except the
``anomalous'' ones \cite{Footnote2} calculated at given chemical potential $\mu$. This approach reduces the problem of
self-consistent diagrammatic calculations of $\delta\Omega$ to a simpler problem of non-self-consistent calculations of
the diagrammatic series $\delta\Omega_\mathrm{BG}$.

\begin{figure}[t]
\begin{center}
\resizebox{0.8\columnwidth}{!}{\includegraphics{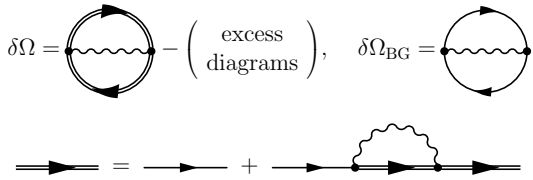}}
\end{center}
\caption{\label{Fig2}The diagrams presenting the interaction-induced correction $\delta\Omega$ to the thermodynamic
potential and its Brueckner-Goldstone part $\delta\Omega_\mathrm{BG}$, which contains no ``anomalous'' diagrams, in the
Hartree-Fock approximation. Thin and thick straight lines are bare and dressed electron Green functions, wiggly line is
the Coulomb interaction.}
\end{figure}

\begin{figure}[t]
\begin{center}
\resizebox{0.8\columnwidth}{!}{\includegraphics{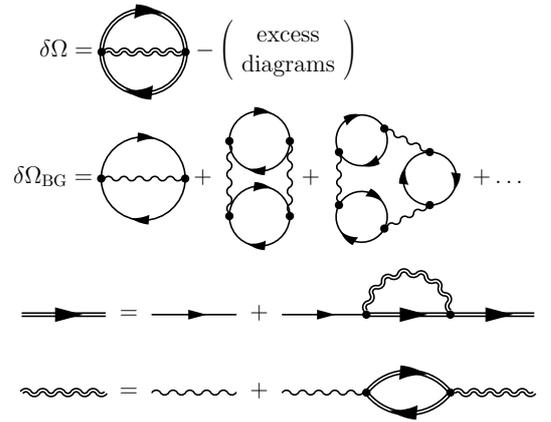}}
\end{center}
\caption{\label{Fig3}The same as Fig.~\ref{Fig2}, but in the self-consistent random-phase approximation. Thick wiggly
line is the screened Coulomb interaction.}
\end{figure}

We apply this approach to the Dirac electron gas in graphene to calculate $E$ in the Hartree-Fock and self-consistent
random-phase approximations. In the Hartree-Fock approximation, the self-consistently evaluated thermodynamic potential
(Fig.~\ref{Fig2}) can be calculated using (\ref{BG}), where the Brueckner-Goldstone perturbation series is reduced to a
single first-order diagram. In the self-consistent random-phase approximation (Fig.~\ref{Fig3}), we take into account
all possible diagrams without vertex corrections. The corresponding Brueckner-Goldstone perturbation series is
presented by the sum of non-self-consistent random-phase approximation diagrams that can be calculated analytically
\cite{Barlas}.

Having analytical results for $\delta\Omega_\mathrm{BG}$ in both approximations (see \cite{Hwang,Barlas} and also
\cite{Lozovik}), we can calculate the ground state energy (\ref{BG}) and substitute it into GVT (\ref{virial_Dirac1}).
Following the lines of Appendix \ref{Appendix_A}, we omit the large constant term (\ref{mu_0_graphene}) in the chemical
potential, which appears due to exchange with the filled valence band and thus actually work with the regularized
energy (\ref{E_reg}) and GVT (\ref{virial_Dirac4}).

In our calculations, we use the following parameters: the bare Fermi velocity is
$v_\mathrm{F}=0.9\times10^6\,\mbox{m/s}$, as retrieved from the comparison of theory and experimental data on graphene
quantum capacitance \cite{Lozovik}. For the cutoff momentum, we assume the value
$p_\mathrm{c}/\hbar=1.095\,\mbox{\AA}^{-1}$, found by equating the density of valence-band electrons $2/S_0$ to the
same density in the Dirac model $gp_\mathrm{c}^2/4\pi\hbar^2$, where $S_0=a^2\sqrt3/2$ is the area of graphene
elementary cell and $g=4$ is the degeneracy factor. For the dielectric constant $\varepsilon$ we take $\varepsilon=1$
(suspended graphene), $\varepsilon=4.5$ (encapsulation in hBN), and $\varepsilon=8$ (some medium with stronger
screening).

\begin{figure}[t]
\begin{center}
\resizebox{1.0\columnwidth}{!}{\includegraphics{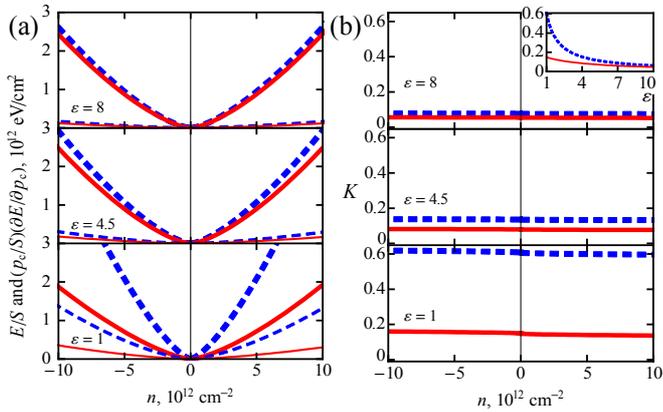}}
\end{center}
\caption{\label{Fig4}(Color online) (a) Ground state energy of electron gas in graphene per unit area $E/S$ (thick
lines) and its derivative $(p_\mathrm{c}/S)(\partial E/\partial p_\mathrm{c})$ (thin lines), calculated as functions of
carrier density $n$ at different values of $\varepsilon$. (b) The normalized derivative $K=(p_\mathrm{c}/E_0)(\partial
E/\partial p_\mathrm{c})$, as defined by (\ref{K}). All curves present calculations in the Hartree-Fock (dashed lines)
and self-consistent random-phase (solid line) approximations. Inset: $K$ at $n\rightarrow0$ as a function of the
dielectric constant $\varepsilon$.}
\end{figure}

Our results for uniform Dirac electron gas of the density $n=N/S$ are presented in Fig.~\ref{Fig4}(a). The first term
in GVT (\ref{virial_Dirac1}), i.e. the ground state energy $E$, is plotted with thick lines. In the absence of both
scale-invariance breaking factors, namely, the system boundary and the momentum cutoff, we would obtain $E=0$. Thus the
nonzero value of $E$ demonstrates the role of these factors. Adding the boundary term, we get the quantity
$E+R(\partial E/\partial R)=p_\mathrm{c}(\partial E/\partial p_\mathrm{c})$, which would be zero in the absence of
momentum cutoff, if conventional VT would have been applicable to graphene. Thus nonzero value of
$p_\mathrm{c}(\partial E/\partial p_\mathrm{c})$ [shown in Fig.~\ref{Fig4}(a) with thin lines] demonstrates the degree
of violation of conventional VT in graphene due to the momentum cutoff.

As we can see, both approximations show similar results at large $\varepsilon$ and disagree at
$\varepsilon\rightarrow1$. The Hartree-Fock approximation always predicts stronger influence of the scale-invariance
breaking terms. At moderate dielectric constants $\varepsilon\gtrsim2$, the momentum cutoff provides much smaller
contribution to GVT than the boundary of the system, as demonstrated in Fig.~\ref{Fig4}(a) where the thick curves pass
much higher than the thin ones.

\section{Cutoff-induced term in the case of graphene}

Consider the ratio
\begin{eqnarray}
K=\frac{p_\mathrm{c}}{E_0}\frac{\partial E}{\partial p_\mathrm{c}}\label{K}
\end{eqnarray}
of the cutoff-induced term in GVT (\ref{virial_Dirac1}) to the energy
\begin{eqnarray}
E_0=\frac43\sqrt{\frac\pi{g}}S\hbar v_\mathrm{F}|n|^{3/2}\label{E0}
\end{eqnarray}
of the noninteracting Dirac electron gas. Explicit calculations of the ground state energy $E$ provide the following
analytical forms of this ratio:
\begin{eqnarray}
K=\frac{r_\mathrm{s}}4+\mathcal{O}\left(\frac{p_\mathrm{F}}{p_\mathrm{c}}\right)\label{en_der_HF}
\end{eqnarray}
in the Hartree-Fock approximation and
\begin{eqnarray}
K=\frac{r_\mathrm{s}}4+\frac3{2\pi
g}a(gr_\mathrm{s})+\mathcal{O}\left(\frac{p_\mathrm{F}}{p_\mathrm{c}}\right)\label{en_der_scRPA}
\end{eqnarray}
in the self-consistent random-phase approximation. Here $r_\mathrm{s}=e^2/\varepsilon\hbar v_\mathrm{F}$ is the
effective fine-structure constant for graphene, $p_\mathrm{F}=\hbar\sqrt{4\pi|n|/g}$ is the Fermi momentum; the
function $a(gr_\mathrm{s})$ was described in \cite{Lozovik}.

The formulas (\ref{en_der_HF})--(\ref{en_der_scRPA}) show that the cutoff-induced term $p_\mathrm{c}(\partial
E/\partial p_\mathrm{c})$ in (\ref{virial_Dirac1}) becomes proportional to $E_0$ in the limit
$p_\mathrm{c}\rightarrow\infty$ or $|n|\rightarrow0$. In the range of carrier densities
$|n|\lesssim10^{13}\,\mbox{cm}^{-2}$, which is accessible in experiments using the electric field effect, their ratio
$K$ indeed weakly depends on $n$, as shown in Fig.~\ref{Fig4}(b). The magnitude of $K$, ranging from 5 to 20\% at
typical values of $\varepsilon$ [see the inset in Fig.~\ref{Fig4}(b)], demonstrates the relative contribution of the
scale-invariance breaking momentum cutoff to GVT in the case of graphene.

Moreover, the quantity $K$ can be related directly to observable characteristics of the electron gas. Differentiating
GVT in the form (\ref{virial_Dirac3}) at $D=2$ with respect to $N$, assuming $\partial K/\partial N\approx0$ and using
(\ref{E0}), we get:
\begin{eqnarray}
2n^2\frac{d\mu}{dn}-\mu n=\frac32\frac{E_0}{S}K,
\end{eqnarray}
where $\partial E/\partial N=\mu$ and $K$ can be taken at $|n|\rightarrow0$. Since $E_0\propto|n|^{3/2}$ according to
(\ref{E0}), we get:
\begin{eqnarray}
\lim_{|n|\rightarrow0}\frac{2n^2(d\mu/dn)-\mu n}{|n|^{3/2}}=\sqrt{\frac{4\pi}g}\hbar v_\mathrm{F}K.
\end{eqnarray}
From this equation, using again (\ref{E0}), we can restore the cutoff-induced term:
\begin{eqnarray}
p_\mathrm{c}\frac{\partial E}{\partial
p_\mathrm{c}}\approx\frac23S|n|^{3/2}\times\lim_{|n|\rightarrow0}\frac{2n^2(d\mu/dn)-\mu n}{|n|^{3/2}}.\label{K_exp}
\end{eqnarray}
The quantities in the right hand side of this formula can be evaluated from experimentally observable characteristics:
$d\mu/dn$ is related to compressibility and quantum capacitance (see \cite{Lozovik} and references therein), while
$\mu$ can be found by integrating $d\mu/dn$. Thus the cutoff-induced term in GVT (\ref{virial_Dirac1}) can be both
extracted from experimental data [using (\ref{K_exp})] and calculated theoretically [using diagrammatic approach or the
approximate formulas (\ref{en_der_HF})--(\ref{en_der_scRPA})].

\section{Conclusions}

We have addressed the problem of quantum-mechanical many-body virial theorem for electrons in a periodic crystal
lattice. First, we derive VT, which is applicable for electrons described by an arbitrary tight-binding model
(\ref{virial_tb1}). Then we proceed to the model of massless Dirac particles with linear dispersion and a momentum
cutoff imposed at the bottom of the valence band, and derive GVT (\ref{virial_Dirac1}) for this system as well as its
grand canonical counterpart (\ref{virial_Dirac2}). GVT for massless Dirac electrons includes the additional term
$p_\mathrm{c}(\partial E/\partial p_\mathrm{c})$, which is absent in the case of usual electron gas. The equations
(\ref{virial_tb1}), (\ref{virial_Dirac1}), and (\ref{virial_Dirac2}) are the main results of our paper.

Alternative ways to obtain GVT, which start immediately from the Dirac model, were demonstrated. We considered the
Dirac particles with the momentum cutoff $|\mathbf{p}|<p_\mathrm{c}$ imposed in order to bound the system energy from
below. In this case GVT can be obtained by means of restricted minimization (\ref{min1})--(\ref{min2}) of the
variational energy with respect to dilatations of the ground state wave function. Another way is to add an additional
term (\ref{Delta_H}) to the Dirac Hamiltonian, which expels electrons from the states with momenta
$|\mathbf{p}|>p_\mathrm{c}$. Note that GVT can also be deduced by means of diagrammatic or dimensional analysis, that
will be considered elsewhere.

We analyzed GVT on the example of massless electrons in graphene, where the ground state energy was calculated by means
of diagrammatic perturbation series. Using the method of Luttinger and Ward \cite{Luttinger}, we calculated the ground
state energy in the Hartree-Fock and self-consistent random-phase approximations. Our analysis shows that the relative
contribution of the cutoff-induced term $p_\mathrm{c}(\partial E/\partial p_\mathrm{c})$, which enters GVT, is about
5--20\% in the case of graphene. Note that our results for GVT for Dirac materisls are cardinally different from those
in \cite{Stokes}, where independence of $r_\mathrm{s}$ in graphene on electron density was not properly taken into
account.

The obtained GVT demonstrates the role of two physical factors, which break the scale invariance of the system: its
finite boundary and momentum cutoff. These factors restrict electron motion in coordinate and momentum spaces and
manifest themselves through corresponding terms in GVT. In the case of graphene, GVT can be verified experimentally as
described in Sec.~VI. Possible applications of the obtained theorem can include checking the validity of
``relativistic'' density functional calculations of ground state properties of disordered graphene
\cite{Rossi,Polini,Rodriguez-Vega}. In addition, the tight-binding version of VT or its hypervirial generalizations can
be applied to study the properties of interacting electrons in other crystals.

\section*{Acknowledgements}

The work of A.D.Z. was supported by RFBR, and the work of Y.E.L. and A.A.S. was supported by the HSE Basic Research
Program.

\appendix

\section{Chemical potential at Dirac point $\mu_0$}
\label{Appendix_A}

First, let us take the derivative of (\ref{virial_Dirac3}) with respect to $N$ at $N=0$:
\begin{eqnarray}
\left.\left(-\frac{\partial E}{\partial N}+p_\mathrm{c}\frac{\partial^2E}{\partial N\partial p_\mathrm{c}}
\right)\right|_{N=0}=0.
\end{eqnarray}
Changing the order of derivatives and using (\ref{mu_0}), we get
\begin{eqnarray}
p_\mathrm{c}\frac{\partial\mu_0}{\partial p_\mathrm{c}}=\mu_0\label{mu_0_der}
\end{eqnarray}
and thus $\mu_0\propto p_\mathrm{c}$.

Then we can regularize the energy of the Dirac electron gas in the following way:
\begin{eqnarray}
E_\mathrm{reg}=E-\mu_0N.\label{E_reg}
\end{eqnarray}
According to (\ref{mu_0}),
\begin{eqnarray}
\left.\frac{\partial E_\mathrm{reg}}{\partial N}\right|_{N=0}=0.\label{mu_reg_0}
\end{eqnarray}
Substituting (\ref{E_reg}) to (\ref{virial_Dirac3}), we get GVT in terms of the regularized energy:
\begin{eqnarray}
-(D+1)E_\mathrm{reg}+DN\frac{\partial E_\mathrm{reg}}{\partial N}+p_\mathrm{c}\frac{\partial E_\mathrm{reg}}{\partial
p_\mathrm{c}}=0,\label{virial_Dirac4}
\end{eqnarray}
which has the same form as (\ref{virial_Dirac3}) due to the relation (\ref{mu_0_der}).

Finally, using the Legendre transformation
\begin{eqnarray}
E_\mathrm{reg}(N,p_\mathrm{c})=\Omega_\mathrm{reg}(\mu_\mathrm{reg},p_\mathrm{c})+\mu_\mathrm{reg}N
\end{eqnarray}
we can convert (\ref{virial_Dirac4}) to GVT in terms of the regularized thermodynamic potential:
\begin{eqnarray}
-(D+1)\Omega_\mathrm{reg}+\mu_\mathrm{reg}\frac{\partial
\Omega_\mathrm{reg}}{\partial\mu_\mathrm{reg}}+p_\mathrm{c}\frac{\partial \Omega_\mathrm{reg}}{\partial
p_\mathrm{c}}=0,\label{virial_Dirac5}
\end{eqnarray}
where
\begin{eqnarray}
\mu_\mathrm{reg}=\frac{\partial E_\mathrm{reg}}{\partial N}
\end{eqnarray}
is the regularized chemical potential. According to (\ref{mu_reg_0}), $\mu_\mathrm{reg}=0$ at $N=0$, as it should be
when the chemical potential is measured from the Dirac point.

\section{Operator of momentum cutoff $P_{p_\mathrm{c}}$}
\label{Appendix_B}

Action of $P_{p_\mathrm{c}}$ on a many-body wave function can be represented as:
\begin{eqnarray}
P_{p_\mathrm{c}}\Psi(\mathbf{r}_1\ldots\mathbf{r}_N)=\int
d\mathbf{r}_1'\ldots\mathbf{r}_N'K_{p_\mathrm{c}}(\mathbf{r}_1-\mathbf{r}_1')\\
\times K_{p_\mathrm{c}}(\mathbf{r}_2-\mathbf{r}_2')\ldots K_{p_\mathrm{c}}(\mathbf{r}_N-\mathbf{r}_N')
\Psi(\mathbf{r}_1'\ldots\mathbf{r}_N').\label{P_pc}
\end{eqnarray}
An integral operator with the kernel
\begin{eqnarray}
K_{p_\mathrm{c}}(\mathbf{r})=\int\limits_{|\mathbf{p}|<p_\mathrm{c}}\frac{d\mathbf{p}}{(2\pi\hbar)^D}\,
e^{\frac{i}\hbar\mathbf{pr}}
\end{eqnarray}
projects a one-particle wave function onto a subspace of momenta $|\mathbf{p}|<p_\mathrm{c}$; this kernel has the scale
property:
\begin{eqnarray}
K_{p_\mathrm{c}}(\lambda\mathbf{r})=\frac1{\lambda^D}K_{\lambda p_\mathrm{c}}(\mathbf{r}).
\end{eqnarray}
Using it in (\ref{P_pc}), we get the following scale property of $P_{p_\mathrm{c}}$:
\begin{eqnarray}
D_\lambda^{-1}P_{p_\mathrm{c}}D_\lambda=P_{\lambda p_\mathrm{c}}.
\end{eqnarray}
For infinitesimal dilatations ($\lambda\rightarrow1$) we get, according to (\ref{D}), that:
\begin{eqnarray}
[P_{p_\mathrm{c}},G]=i\hbar p_\mathrm{c}\frac{\partial P_{p_\mathrm{c}}}{\partial p_\mathrm{c}}.\label{P_pc_comm}
\end{eqnarray}


\begin{thebibliography}{99}
\bibitem{Footnote1}
We omit the energy $\langle\Psi|H_\mathrm{ext}|\Psi\rangle$ of electron interaction with the external potential because
it vanishes in the bulk limit or when $U_\mathrm{ext}(\mathbf{r})$ is approximated by potential walls, which height
tends to infinity.
\end{thebibliography}
\end{document}